\documentclass[american,english]{article}
\usepackage[]{fontenc}
\usepackage[latin1]{inputenc}
\usepackage{amsmath}
\usepackage{graphicx}
\usepackage{amssymb}

\makeatletter

\usepackage{babel}
\makeatother
\begin{document}

\title{Vector current conservation and neutrino emission from singlet-paired
baryons in neutron stars.}

\author{L. B. Leinson$^{1,2}$ and A. P\'{e}rez$^{2}$\\
$^{1}$Institute of Terrestrial Magnetism, Ionosphere and\\
 Radio Wave Propagation\\
 RAS, 142190 Troitsk, Moscow Region, Russia\\
 $^{2}$Departamento de F\'{\i}sica Te\'{o}rica and IFIC, \\
 Universidad de Valencia-CSIC, \\
 Dr. Moliner 50, 46100--Burjassot, Valencia, Spain}

\maketitle
\begin{abstract}
Neutrino emission caused by singlet Cooper pairing of baryons in neutron
stars is recalculated by accurately taking into account for conservation
of the vector weak currents. The neutrino emissivity via the vector
weak currents is found to be several orders of magnitude smaller than
that obtained before by different authors. This makes unimportant
the neutrino radiation from singlet pairing of protons or hyperons. 
\end{abstract}
One of the mechanisms leading the neutron star cooling, specially
for temperatures near the critical one $T_{c}$, consists on the recombination
of thermally excited baryon BCS pairs into the condensate. This process
has received the attention of many authors, and is currently thought
to be dominant, for some ranges of the temperature and/or matter density
(modulo the uncertainties arising from the incomplete knowledge of
the gap value). A better understanding of this process is, therefore,
of great importance for the secular evolution of such objects. 

Under the description of nuclear matter in the nonrelativistic limit,
the simplest case for baryon pairing corresponds to two particles
bounded in the $^{1}S_{0}$ state. The neutrino emission for recombination
into this state was first calculated by Flowers et al. \cite{FRS76}
and reproduced by other authors \cite{Vosk}, \cite{YKL98}. The neutrino
energy losses due to pairing of hyperons \cite{Balberg}, \cite{Schaab}
are also discussed in the literature as possible cooling mechanisms
for superdense baryonic matter in neutron stars. Nowadays, these ideas
are widely accepted and used in numerical simulations of neutron star
evolution \cite{Sch}, \cite{Page98}, \cite{Yak98}. 

In the case of singlet pairing, the averaged weak axial current vanishes,
and the emissivity is directly mediated by the weak vector current.
As it is well known, the vector current possesses the property of
being conserved by nuclear and electroweak interactions. Of course,
this fundamental property has to be accounted for in any calculation
of neutrino emission processes. As we show below, however, previous
calculations did not pay attention to this particular topic. This
translates into a dramatic overestimation of the energy production
from the process under consideration.

Let us recall shortly the main steps in the above calculations. The
low-energy Lagrangian of the weak interaction may be described by
a point-like current-current approach. For interactions mediated by
neutral weak currents, it can be written as%
\footnote{In what follows we use the Standard Model of weak interactions, the
system of units $\hbar=c=1$ and the Boltzmann constant $k_{B}=1$.
The fine-structure constant is $\alpha=e^{2}/4\pi=1/137$.%
} \begin{equation}
\mathcal{L}_{vac}=\frac{G_{F}}{2\sqrt{2}}J_{B}^{\mu}l_{\mu}.\end{equation}
Here $G_{F}$ is the Fermi coupling constant, and the neutrino weak
current is given by $l_{\mu}=\bar{\nu}\gamma_{\mu}\left(1-\gamma_{5}\right)\nu$.
The vacuum weak current of the baryon is of the form $J_{\mu}=\bar{\psi}\left(C_{V}\gamma_{\mu}-C_{A}\gamma_{\mu}\gamma_{5}\right)\psi$,
where $\psi$ represents the baryon field, and the weak vertex includes
the vector and axial-vector terms with the corresponding coupling
constants $C_{V}$ and $C_{A}$.

As mentioned above, in the case of singlet pairing only the vector
current contributes. The nonrelativistic limit for this current is
taken as $\bar{\psi}_{B}\gamma^{0}\psi_{B}\rightarrow\hat{\Psi}_{B}^{+}\hat{\Psi}_{B}$,
all others being zero. Here $\hat{\Psi}_{B}$ is the second-quantized
nonrelativistic spinor wave function. The process is kinematically
allowed due to the existence of a superfluid energy gap $\Delta$,
which admits the transition with time-like momentum transfer $K=\left(\omega,\mathbf{k}\right)$,
with $\omega=\omega_{1}+\omega_{2}$ and $\mathbf{k=k}_{1}+\mathbf{k}_{2}$
being the energy and momentum carried out by the freely escaping neutrino
pair. We have $\omega>2\Delta$ and $\omega>k$ .

The relevant input for this calculation is the recombination matrix
element between the baryon state, which has a pair of quasi-particle
excitations of momentum-spin labels $\left(\mathbf{p},\mathrm{up};\mathbf{p}^{\prime},\mathrm{down}\right)$,
and the same state but with these excitations restored to the condensate.
To the leading (zero) order in $k\ll p_{F}$, this matrix element
is usually estimated as \cite{FRS76}\begin{equation}
\left\vert \mathcal{M}_{B}\right\vert ^{2}=\frac{\Delta^{2}}{\epsilon_{p}^{2}}\label{ME}\end{equation}
where $\epsilon_{p}$ is the quasi-particle energy, as given by Eq.
(\ref{Ep}). As a result, the neutrino energy losses at temperature
$T<T_{c}$ are found to be:\begin{equation}
Q_{\mathrm{FRS}}=\frac{4G_{F}^{2}{p}_{F}M^{\ast}C_{V}^{2}}{15\pi^{5}}\mathcal{N}_{\nu}T^{7}y^{2}\int_{0}^{\infty}\;\frac{z^{4}dx}{\left(e^{z}+1\right)^{2}},\label{SF}\end{equation}
where $M^{\ast}$ is the effective nucleon mass, $y=\Delta/T$, $z=\sqrt{x^{2}+y^{2}}$,
and $\mathcal{N}_{\nu}=3$ is the number of neutrino flavors.

The naive estimate (\ref{ME}) is inconsistent with the hypothesis
of conservation of the vector current in weak interactions. Indeed,
a longitudinal vector current of quasi-particles consisting only on
a temporal component can not satisfy the continuity equation. It is
well known, however, that the Bardeen-Cooper-Schrieffer theory of
superconductivity is gauge invariant \cite{many} and that the current
conservation can be restored if the interaction among quasi-particles
is incorporated in the coupling vertex to the same degree of approximation
as the self-energy effect is included in the quasi-particle \cite{Bogoliubov},
\cite{Nambu}. In the present paper we recalculate the neutrino energy
losses with allowance for conservation of the weak vector current.

It is convenient to use the Nambu-Gorkov formalism, where the quasi-particle
fields are represented by two-component objects\begin{equation}
\Psi\left(p\right)=\left(\begin{array}{c}
\psi_{\mathbf{1}}\left(p\right)\\
\psi_{2}^{\dagger}\left(-p\right)\end{array}\right).\label{tcf}\end{equation}
Here $\psi_{\mathbf{1}}\left(p\right)$ is the the quasi-particle
component of the excitation with momentum $\mathbf{p}$ and spin $\sigma$,
and $\psi_{2}^{\dagger}\left(-p\right)$ is the hole component of
the same excitation, which can be interpreted as the absence of a
particle with momentum $-\mathbf{p}$ and spin $-\sigma$. The two-component
fields (\ref{tcf}) obey the standard fermion commutation relations\[
\left\{ \Psi_{p,\sigma},\Psi_{p^{\prime},\sigma^{\prime}}\right\} =\delta_{\sigma,\sigma^{\prime}}\delta_{p,p^{\prime}}.\]
With the aid of the $2\times2$ Pauli matrices $\hat{\tau}_{i}=\left(\hat{\tau}_{1},\hat{\tau}_{2},\hat{\tau}_{3}\right)$
operating in the particle-hole space, the Hamiltonian of the system
of quasi-particles can be recast as \cite{Nambu}\[
H=H_{0}+H_{1},\]
where \begin{equation}
H_{0}=\sum_{p}\Psi_{p}^{\dagger}\left(\xi_{\mathbf{p}}\hat{\tau}_{3}+\Delta\hat{\tau}_{1}\right)\Psi_{p}\label{Ham}\end{equation}
is the BCS reduced Hamiltonian, the nonrelativistic energy is measured
relatively to the Fermi level\[
\xi_{\mathbf{p}}\equiv\frac{p^{2}}{2M^{\ast}}-\mu,\]
and $\mu$ is the Fermi energy. The BCS reduced Hamiltonian (\ref{Ham})
bears a resemblance to the one describing the Dirac equation. It has
eigenvalues $p_{0}=\pm\epsilon_{p}$ with\begin{equation}
\ \epsilon_{p}=\sqrt{\xi_{\mathbf{p}}^{2}+\Delta^{2}},\label{Ep}\end{equation}
which correspond to excited states in the particle-hole picture, while
the ground state (vacuum) is the state where all negative energy \char`\"{}quasi-particles\char`\"{}
($\epsilon<0$) are occupied and no positive energy particles exist.
The positive and negative states are separated by an energy gap $2\Delta$.

The Hamiltonian of residual interaction among quasi-particles has
the following form \[
H_{1}=\frac{1}{2}\sum_{p^{\prime},q}V_{pp^{\prime}}\left(q\right)\left(\Psi_{p+q}^{\dagger}\hat{\tau}_{3}\Psi_{p}\right)\left(\Psi_{p^{\prime}-q}^{\dagger}\hat{\tau}_{3}\Psi_{p\prime}\right).\]
As follows from the Hamiltonian (\ref{Ham}), the inverse of the quasi-particle
propagator can be written as \cite{Nambu}:\begin{equation}
\text{\c{G}}^{-1}\ =\ p_{0}-\xi_{\mathbf{p}}\hat{\tau}_{3}-\Delta\hat{\tau}_{1},\label{Gn}\end{equation}

The self-energies are, in general, complex numbers due to the instability
of single particles. However, to the extent that the single-particle
picture makes some physical sense, we will ignore the small imaginary
part of the self-energies, and describe the quasi-particles with the
aid of wave-functions. The states of quasi-particles obey the equation\begin{equation}
\text{\c{G}}^{-1}\Psi_{\mathbf{p}}=0.\label{e}\end{equation}
The solution to this equation corresponding to the energy $p_{0}=\epsilon_{\mathbf{p}}$
and spin state $\chi_{\sigma}$ has the following form\begin{equation}
\Psi_{\mathbf{p},\sigma}=\left(\begin{array}{c}
u_{\mathbf{p}}\chi_{\sigma}\\
v_{\mathbf{p}}\chi_{-\sigma}\end{array}\right)e^{i\mathbf{pr}-i\epsilon_{\mathbf{p}}t}\label{vfp}\end{equation}
with\[
u_{\mathbf{p}}=\sqrt{\frac{\epsilon_{p}+\xi_{p}}{2\epsilon_{p}}},\ \ \ \
v_{\mathbf{p}}=\sqrt{\frac{\epsilon_{p}-\xi_{p}}{2\epsilon_{p}}}.\]
There is also a solution of negative frequency $p_{0}=-\epsilon_{\mathbf{p}}$\begin{equation}
\Psi_{-\mathbf{p},-\sigma}=\left(\begin{array}{c}
-v_{\mathbf{p}}\chi_{-\sigma}\\
u_{\mathbf{p}}\chi_{\sigma}\end{array}\right)e^{-i\mathbf{pr+}i\epsilon_{\mathbf{p}}t},\label{vfm}\end{equation}
which corresponds to the same excitation energy. This solution is
connected to the hole state by the particle-antiparticle conjugation\[
C:\ \Psi^{C}=C\Psi^{\dagger}=\hat{\tau}_{2}\Psi^{\dagger}.\]
which changes quasi-particles of energy-momentum $\left(p_{0},\mathbf{p}\right)$
into holes of energy-momentum $\left(-p_{0},-\mathbf{p}\right)$.

The components of the bare vertex \begin{equation}
\gamma^{\mu}=\left\{ \begin{array}{cc}
\hat{\tau}_{3} & \mathrm{if\ }\mu=0,\ \ \ \ \ \ \ \ \ \ \ \ \\
\frac{1}{M^{\ast}}\mathbf{p} & \mathrm{if\ }\mu=i=1,2,3\end{array}\right..\label{gam}\end{equation}
are $2\times2$ matrices in the Nambu-Gorkov space. As already mentioned,
the longitudinal current corresponding to the bare vertex does not
satisfy the continuity equation. To restore the current conservation,
one must consider the modification of the vertex $\gamma^{\mu}$ to
the same order as the modification of the propagator is done. The
relation between the modified vertex $\Gamma^{\mu}$ and the quasi-particle
propagator (\ref{Gn}) is given by the Ward identity \cite{Schr}
\begin{equation}
K_{\mu}\Gamma^{\mu}\left(p^{\prime},p\right)=\hat{\tau}_{3}\text{\c{G}}^{-1}\left(p\right)-\text{\c{G}}^{-1}\left(p^{\prime}\right)\hat{\tau}_{3},\label{Ward}\end{equation}
where $K=\left(\omega,\mathbf{k}\right)$ is the transferred momentum.
The plane wave solutions \[
u_{\mathbf{p},\alpha}\exp\left(i\mathbf{pr}-i\epsilon_{\mathbf{p}}t\right),\ \ \ \ \ u_{\mathbf{p}^{\prime},\alpha^{\prime}}^{\ast}\exp\left(-i\mathbf{pr}+i\epsilon_{\mathbf{p}}t\right)\]
for $\Psi$ and $\Psi^{+}$ obey the equations \c{G}$^{-1}\left(p\right)u_{\mathbf{p},\alpha}=0$,
and $u_{\mathbf{p}^{\prime},\alpha^{\prime}}^{\ast}$\c{G}$^{-1}\left(p^{\prime}\right)=0$.
Therefore, the Ward identity implies conservation of the vector current
on the energy shell of the quasi-particles. Following the prescriptions
of quantum electrodynamics, an approximation which satisfies the Ward
identity (and hence the continuity equation) is the sum of ladder
diagrams.

Consider first the case of electrically neutral baryons. Then the
corrected vertex can be found from the following Dyson equation\begin{eqnarray}
\Gamma^{\mu}\left(p-K,p\right) & = & \hat{\gamma}^{\mu}\left(p-K,p\right)\label{Gam}\\
 &  & +i\int\frac{d^{4}p^{\prime}}{\left(2\pi\right)^{4}}\ \hat{\tau}_{3}\text{\c{G}}\left(p^{\prime}-K\right)\Gamma^{\mu}\left(p^{\prime}-K,p^{\prime}\right)\text{\c{G}}\left(p^{\prime}\right)\hat{\tau}_{3}V_{pp^{\prime}},\notag\end{eqnarray}
where the \char`\"{}dressed\char`\"{} particles interact with the
same primary interaction $V_{pp^{\prime}}$ which produces the self-energy
of the quasi-particle.

In the limit $K=\left(\omega,\mathbf{0}\right)$, the Ward identity
gives%
\footnote{To obtain the weak vector current this vertex should be multiplied
by the weak coupling constant $C_{V}$.%
} \[
\Gamma^{0}\left(p-K,p\right)=\hat{\tau}_{3}-\frac{2}{\omega}i\hat{\tau}_{2}\Delta\]
The poles of the vertex function correspond to collective eigen-modes
of the system. Therefore, the pole which appears at $\omega\rightarrow0$,
$k=0$ implies the existence of a collective mode, which plays an
important role in the conservation of the vector current. The corresponding
nonperturbative solution to Eq. (\ref{Gam}) has been found by Nambu
\cite{Nambu} (see also \cite{Littlewood}): \begin{equation}
\Gamma_{0}\left(p-K,p\right)=\hat{\tau}_{3}-2i\hat{\tau}_{2}\Delta\frac{\omega}{\omega^{2}-a^{2}k^{2}}\label{G0}\end{equation}
\begin{equation}
\mathbf{\Gamma}=\frac{\mathbf{p}}{M}-2i\hat{\tau}_{2}\Delta\frac{a^{2}\mathbf{k}}{\omega^{2}-a^{2}k^{2}},\label{GV}\end{equation}
The poles in this vertex correspond to the collective motion of the
condensate, with the dispersion relation $\omega=ak$, where $a^{2}=V_{F}^{2}/3$.

The effective vertex satisfies the Ward identity (\ref{Ward}), and
thus the continuity equation on the energy shell\begin{equation}
\omega\Gamma_{0}-\mathbf{k\Gamma}\simeq0\label{cvc}\end{equation}

We are now in a position to evaluate the matrix element of the vector
weak current. In the particle-hole picture, the creation and recombination
of two quasi-particles is described by the off-diagonal matrix elements
of the Hamiltonian, which corresponds to quasi-particle transitions
into a hole (and a correlated pair). Thus, we calculate the matrix
element of the current between the initial (positive-frequency) state
of a quasi-particle with momentum $\mathbf{p}$ and the final (negative-frequency)
state with the same momentum $\mathbf{p}$. \[
\mathcal{M}_{\mu}=\left\langle \Psi_{-\mathbf{p,}-\sigma}^{\dagger}\left\vert \Gamma_{\mu}\right\vert \Psi_{\mathbf{p,\sigma}}\right\rangle \]

Let us consider separately the contributions from the bare vertex,
given by the first term in Eq. (\ref{G0}), and the collective part,
given by the second term, so that $\mathcal{M}_{\mu}=\mathcal{M}_{\mu}^{\mathrm{bare}}+\mathcal{M}_{\mu}^{\mathrm{coll}}$.

Making use of the wave functions described by Eqs. (\ref{vfp}), (\ref{vfm})
for $\mu=0$, we find \begin{equation}
\mathcal{M}_{0}^{\mathrm{bare}}=-\left(u_{p}v_{p^{\prime}}+v_{p}u_{p^{\prime}}\right)\simeq-\frac{\Delta_{\mathbf{\check{p}}}}{\epsilon_{p}},\ \ \ \ k\ll p\simeq p_{F}\label{first}\end{equation}
\[
\mathcal{M}_{0}^{\mathrm{coll}}=2\Delta_{\mathbf{\check{p}}}\frac{\omega}{\omega^{2}-a^{2}k^{2}}\left(u_{p}u_{p^{\prime}}+v_{p}v_{p^{\prime}}\right)\simeq2\Delta_{\mathbf{\check{p}}}\frac{\omega}{\omega^{2}-a^{2}k^{2}}.\]
with $\epsilon_{p}+\epsilon_{p^{\prime}}=\omega$ and $\mathbf{p+p}^{\prime}=\mathbf{k}$.

The velocity of the collective mode $a^{2}=V_{F}^{2}/3$ is small
in the nonrelativistic system. Therefore, we expand the collective
contribution in this parameter to obtain\begin{equation}
\mathcal{M}_{0}^{\mathrm{coll}}\simeq\frac{\Delta_{\mathbf{\check{p}}}}{\epsilon_{p}}\left(1+\frac{1}{3}V_{F}^{2}\frac{k^{2}}{\omega^{2}}\right).\label{second}\end{equation}

The contribution of the bare vertex $\mathcal{M}_{0}^{\mathrm{bare}}$
reproduces the matrix element (\ref{ME}) derived by Flowers et al.
\cite{FRS76} and Yakovlev et al. \cite{YKL98}. However, the collective
correction modifies this crucially. In the sum of the two contributions,
the leading terms mutually cancel, yielding the matrix element\[
\mathcal{M}_{0}=\mathcal{M}_{0}^{\mathrm{bare}}+\mathcal{M}_{0}^{\mathrm{coll}}\simeq\frac{1}{3}V_{F}^{2}\frac{k^{2}}{\omega^{2}}\frac{\Delta_{\mathbf{\check{p}}}}{\epsilon_{p}}\]
which is at least $\sim V_{F}^{2}$ times smaller than the bare result.

The spatial component of the longitudinal (with respect to $\mathbf{k}$)
component of the matrix element can be obtained from Eq. (\ref{cvc}).
Since $\mathbf{\check{k}\Gamma=}\left(\omega/k\right)\Gamma_{0}$
we have\[
\mathcal{M}_{\parallel}\simeq\frac{1}{3}V_{F}^{2}\frac{k}{\omega}\frac{\Delta_{\mathbf{\check{p}}}}{\epsilon_{p}}.\]
In the above, $\ \mathbf{\check{k}=k}/k$ is a unit vector directed
along the transferred momentum.

Since the collective interaction modifies only the longitudinal part
of the vertex, the transverse part of the matrix element can be evaluated
directly from the bare vertex (\ref{gam}). This yields\[
\mathcal{M}_{\perp}\simeq\left(v_{p}u_{p^{\prime}}-u_{p}v_{p^{\prime}}\right)\frac{\mathbf{p}_{\perp}}{M^{\ast}}\simeq-\frac{1}{2}V_{F}^{2}\frac{k\Delta_{\mathbf{\check{p}}}}{\epsilon_{\mathbf{p}}^{2}}\left(\mathbf{\check{k}\check{p}}\right)\mathbf{\check{p}}_{\perp}\]
with $\mathbf{\check{p}=p}/p$.

The rate of the process is proportional to the square of the matrix
element. This means that the vector current contribution to the neutrino
energy losses is $V_{F}^{4}$ times smaller than estimated before.
The corresponding neutrino emissivity in the vector channel can be
evaluated with the aid of Fermi's golden rule: \begin{eqnarray*}
Q_{V} & = & \left(\frac{G_{F}}{2\sqrt{2}}\right)^{2}\frac{C_{V}^{2}}{\left(2\pi\right)^{8}}\mathcal{N}_{\nu}\int d^{3}pd^{3}p^{\prime}f\left(\epsilon_{\mathbf{p}}\right)f\left(\epsilon_{\mathbf{p}^{\prime}}\right)\\
 &  & \times\int\frac{d^{3}k_{1}}{2\omega_{1}}\frac{d^{3}k_{2}}{2\omega_{2}}\omega\mathrm{Tr}\left(l_{\mu}l_{\nu}^{\ast}\right)\mathcal{M}^{\mu}\mathcal{M}^{\nu}\delta\left(\mathbf{p+p}^{\prime}-\mathbf{k}\right)\delta\left(\epsilon_{\mathbf{p}}+\epsilon_{\mathbf{p}^{\prime}}-\omega\right).\end{eqnarray*}
One can simplify this equation by inserting $\int d^{4}K\ \delta^{\left(4\right)}\left(K-k_{1}-k_{2}\right)=1$.
Then, the phase-space integrals for neutrinos are readily done with
the aid of Lenard's formula \begin{eqnarray*}
 &  &
 \int\frac{d^{3}k_{1}}{2\omega_{1}}\frac{d^{3}k_{2}}{2\omega_{2}}\delta^{\left(4\right)}\left(K-k_{1}-k_{2}\right)\mathrm{Tr}\left(l^{\mu}l^{\nu\ast}\right)\\
 & = & \frac{4\pi}{3}\left(K_{\mu}K_{\nu}-K^{2}g_{\mu\nu}\right)\Theta\left(K^{2}\right)\Theta\left(\omega\right),\end{eqnarray*}
where $\Theta(x)$ is the Heaviside step function.

For $k\ll p_{F}$ we obtain\begin{eqnarray*}
Q_{V} & = & \frac{4\pi}{3}\left(\frac{G_{F}}{2\sqrt{2}}\right)^{2}\frac{C_{V}^{2}}{\left(2\pi\right)^{8}}\mathcal{N}_{\nu}\int d^{3}pf^{2}\left(\epsilon_{\mathbf{p}}\right)\int_{0}^{\infty}d\omega\omega\int_{0}^{\omega}dkk^{2}d\Omega_{k}\\
 &  & \times\left(\left(K_{\mu}\mathcal{M}^{\mu}\right)^{2}-K^{2}\mathcal{M}^{\mu}\mathcal{M}_{\mu}\right)\delta\left(2\epsilon_{\mathbf{p}}-\omega\right).\end{eqnarray*}
The next integrations are trivial. We get\[
Q_{V}=\frac{592}{42\,525\pi^{5}}V_{F}^{4}G_{F}^{2}C_{V}^{2}p_{F}M^{\ast}T^{7}y^{2}\int_{0}^{\infty}\;\frac{z^{4}dx}{\left(e^{z}+1\right)^{2}}\]
This is to be compared with Eq. (\ref{SF}). We see that \[
\frac{Q_{V}}{Q_{\mathrm{FRS}}\left(_{1}S^{0}\right)}=\frac{148}{2835}V_{F}^{4}\]
i.e. the neutrino radiation from $^{1}S_{0}$ pairing in the nonrelativistic
system $\left(V_{F}\ll1\right)$ is suppressed by several orders of
magnitude with respect to the predictions of \cite{FRS76} and \cite{YKL98}.

Consider now the case when quasi-particles carry an electric charge.
Including the long-range Coulomb interaction $V_{C}\left(k\right)=e^{2}/k^{2}$
implies that the vertex part is multiplied by a string of closed loops,
which represents the polarization of the surrounding medium. In this
case, the new vertex $\tilde{\Gamma}^{\mu}$ can be found as the solution
of the Dyson equation, according to the diagram of Fig. 1 or, analytically
\begin{eqnarray}
\tilde{\Gamma}^{\mu}\left(p-K,p\right) & = & \Gamma^{\mu}\left(p-K,p\right)-\Gamma_{0}\left(p-K,p\right)V_{C}\left(k\right)\notag\\
 &  & \times i\int\frac{d^{4}p^{\prime}}{\left(2\pi\right)^{4}}\ Tr\left[\hat{\tau}_{3}\text{\c{G}}\left(p^{\prime}-K\right)\tilde{\Gamma}^{\mu}\left(p^{\prime}-K,p^{\prime}\right)\text{\c{G}}\left(p^{\prime}\right)\right]\label{gc}\end{eqnarray}

\begin{figure}
\includegraphics{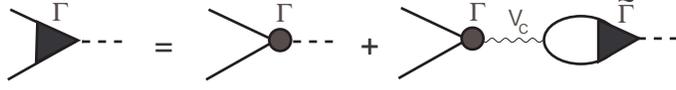}

\caption{Dyson equation for the vertex correction for charged quasi-particles.
The shaded areas represent the modified effective vertex, and the
wavy line stands for the Coulomb interaction. }
\end{figure}
This equation can be readily solved yielding

\[
\tilde{\Gamma}^{\mu}\left(p-K,p\right)=\Gamma^{\mu}\left(p-K,p\right)\left(1-\frac{V_{C}\left(k\right)\Pi^{0\mu}\left(K\right)}{1+V_{C}\left(k\right)\Pi^{00}\left(K\right)}\right).\]
with\[
\Pi^{0\mu}\left(K\right)\equiv i\int\frac{d^{4}p^{\prime}}{\left(2\pi\right)^{4}}\ Tr\left[\hat{\tau}_{3}\text{\c{G}}\left(p^{\prime}-K\right)\Gamma^{\mu}\left(p^{\prime}-K,p^{\prime}\right)\text{\c{G}}\left(p^{\prime}\right)\right].\]
In particular, for $\Gamma^{0}$ we arrive to%
\footnote{The solution to the equation $1+V_{C}\left(k\right)\Pi^{00}\left(K\right)=0$
determines the new dispersion law $\omega=\omega\left(k\right)$ for
the collective excitations, which represents plasma waves \cite{Nambu}.%
}\begin{equation}
\tilde{\Gamma}^{0}\left(p-K,p\right)=\frac{\Gamma^{0}\left(p-K,p\right)}{1+V_{C}\left(k\right)\Pi^{00}\left(K\right)}.\label{G0em}\end{equation}

The polarization function $\Pi^{00}\left(K\right)$ can be readily
calculated with the help of $\Gamma^{0}$ given by Eq. (\ref{G0}).
By neglecting the small dependence of the energy gap on the transferred
momentum $\mathbf{k}$, we have\begin{eqnarray*}
\Pi^{00}\left(K\right) & = & i\int\frac{d^{4}p^{\prime}}{\left(2\pi\right)^{4}}\ Tr\left[\hat{\tau}_{3}\text{\c{G}}\left(p^{\prime}-K\right)\hat{\tau}_{3}\text{\c{G}}\left(p^{\prime}\right)\right]\\
 &  & -\frac{2\omega}{\omega^{2}-a^{2}k^{2}}i\int\frac{d^{4}p^{\prime}}{\left(2\pi\right)^{4}}\ Tr\left[\hat{\tau}_{3}\text{\c{G}}\left(p^{\prime}-K\right)i\hat{\tau}_{2}\Delta\left(p^{\prime}\right)\text{\c{G}}\left(p^{\prime}\right)\right].\end{eqnarray*}
Here, the quasi-particle propagator follows from Eq. (\ref{Gn}):
\begin{equation}
\text{\c{G}}\left(p\right)=\frac{i}{p_{0}^{2}-\epsilon_{p}^{2}}\left(p_{0}+\xi_{p}\hat{\tau}_{3}+\hat{\tau}_{1}\Delta\right).\label{Gp}\end{equation}
We are interested in the regime defined by $k<\omega,\ \ \ \omega>2\Delta\gg\epsilon_{\mathbf{p}}-\epsilon_{\mathbf{p-k}}\simeq kV_{F}$.
In this case we obtain, after some simplifications

\begin{equation}
\tilde{\Gamma}^{0}\left(p-K,p\right)=\frac{\Gamma^{0}\left(p-K,p\right)}{1+\chi\left(K\right)},\label{G0em1}\end{equation}
with\begin{eqnarray}
\chi\left(K\right) & = & \frac{e^{2}}{8\pi^{3}}\left[\frac{a^{2}}{\omega^{2}-a^{2}k^{2}}\int d^{3}p\ \allowbreak\frac{\Delta^{2}}{\epsilon_{p}\left(\epsilon_{p}^{2}-a^{2}k^{2}/4\right)}\right.\notag\\
 &  & \left.+\int d^{3}p\ \frac{\Delta^{2}}{\epsilon_{p}\left(\omega^{2}-4\epsilon_{p}^{2}\right)}\allowbreak\left(\frac{p^{2}}{3M^{2}\epsilon_{p}^{2}}-\frac{a^{2}}{\epsilon_{p}^{2}-a^{2}k^{2}/4}\right)\right].\label{hi}\end{eqnarray}
Since $a\ll1$ and $p\simeq p_{F}$, the second integral in Eq. (\ref{hi})
may be dropped. By neglecting also the small contributions from $a^{2}k^{2}\ll\epsilon_{p}^{2},\omega^{2}$
we get\[
\chi\left(K\right)=e^{2}\frac{a^{2}}{\omega^{2}}\int\ \allowbreak\frac{\Delta^{2}}{\epsilon_{p}^{3}}\frac{d^{3}p}{\left(2\pi\right)^{3}}=\frac{\omega_{p}^{2}}{\omega^{2}}\]
with $\omega_{p}^{2}=e^{2}n/M^{\ast}$ ($n$ is the number of baryons
per unit volume). This agrees with the plasma frequency for a free
gas of charged particles.

The energy exchange in the medium goes naturally as the temperature
scale. Therefore, the energy transferred to the radiated neutrino-pair
is $\omega\sim T\leq T_{c}$, while the plasma frequency $\omega_{p}$
is typically much larger than the critical temperature for Cooper
pairing. For instance, for a number density $n$ of the order of the
nuclear saturation density $n_{0}\simeq0.17$ fm and the effective
mass of the baryon $M^{\ast}$ of the order of the bare nucleon mass,
we obtain $\omega_{p}\sim10\ MeV$, while the critical temperature
for baryon pairing is about $1\ MeV$ or less. Under these conditions,
we obtain

\[
\tilde{\Gamma}^{0}\left(p-K,p\right)\simeq\frac{\omega^{2}}{\omega_{p}^{2}}\Gamma^{0}\left(p-K,p\right)\sim\frac{T_{c}^{2}}{\omega_{p}^{2}}\Gamma^{0}\left(p-K,p\right)\]
Thus, in superconductors, the vector current contribution to the neutrino
radiation is suppressed additionally by a factor $\left(T_{c}^{2}/\omega_{p}^{2}\right)^{2}$:
this is the \textit{plasma screening effect}.

We have considered the problem of conservation of the vector weak
current in the theory of neutrino-pair radiation from Cooper pairing
in neutron stars. The correction to the vector weak vertex is calculated
within the same order of approximation as the quasi-particle propagator
is modified by the pairing interaction in the system. This correction
restores the conservation of the vector weak current in the quasi-particle
transition into a paired state. As a result, in the nonrelativistic
baryon system, the matrix element of the vector current is $V_{F}^{2}$
times smaller than previous estimations. This means that the vector
weak current contribution to neutrino radiation caused by Cooper paring
is $V_{F}^{4}$ times smaller than it was thought before. The vector
weak current contribution from pairing of charged baryons is suppressed
additionally by a factor $\sim\left(T_{c}^{2}/\omega_{p}^{2}\right)^{2}$
due to plasma screening. The total suppression factor due to both
the current conservation and the plasma effects is of the order \[
\left(T_{c}^{2}/\omega_{p}^{2}\right)^{2}V_{F}^{4}\lesssim10^{-6}.\]
Thus the neutrino energy losses due to singlet-state pairing of baryons
can, in practice, be neglected in simulations of neutron star cooling.
This makes unimportant the neutrino radiation from pairing of protons
or hyperons.

\selectlanguage{american}
\textbf{Acknowledgments}

\selectlanguage{english}
This work has been supported by Spanish Grants AYA2004-08067-C01,
FPA2005-00711 and GV2005-264.


\begin{thebibliography}{10}
\bibitem{FRS76} E. Flowers, M. Ruderman, P. Sutherland, ApJ 205 (1976)
541.

\bibitem{Vosk} D. N. Voskresensky, and A. V. Senatorov, Sov. J. Nucl.
Phys. 45 (1987) 411.

\bibitem{YKL98} D. G. Yakovlev, A. D. Kaminker, K. P. Levenfish,
A\&A 343 (1999) 650.

\bibitem{Balberg} S. Balberg, N. Barnea, Phys. Rev. 57C (1998) 409.

\bibitem{Schaab} Ch. Schaab, S. Balberg, J. Schaffner-Bielich, ApJL
(1998) 504, L99.

\bibitem{Sch} C. Shaab, D. Voskresensky, A. D. Sedrakian, F. Weber,
M. K. Weigel A\&A, 321 (1997) \textbf{}591.

\bibitem{Page98} D. Page, In: Many Faces of Neutron Stars (eds. R.
Buccheri, J. van Peredijs, M. A. Alpar. Kluver, Dordrecht, 1998) p.
538.

\bibitem{Yak98} D. G. Yakovlev, A. D. Kaminker, K. P. Levenfish,
In: Neutron Stars and Pulsars (ed. N. Shibazaki et al., Universal
Akademy Press, Tokio, 1998) p. 195.

\bibitem{many} M. J. Buckingam, Nuovo cimento, 5 (1957) 1763; J.
Bardeen, Nuovo cimento, 5 (1957) 1765; M. R Schafroth, Phys. Rev.
111 (1958) 72; P. W. Anderson, Phys. Rev. 110 (1958) 827; 112 (1958)
1900; G. Rickayzen, Phys. Rev. 111 (1958) 817; Phys. Rev. Lett. 2
(1959) 91; D. Pines and R. Schieffer, Nuovo cimento 10 (1958) 496;
Phys. Rev. Lett. 2 (1958) 407; J. M Blatt and T. Matsubara, Progr.
Theor. Phys. 20 (1958) 781.

\bibitem{Bogoliubov} N. N. Bogoliubov, Soviet Phys. 34 (1958) 41,
51.

\bibitem{Nambu} Y. Nambu, Phys. Rev. 117 (1960) 648.

\bibitem{Schr} J. Schrieffer, Theory of Superconductivity (W. Benjamin,
New York, 1964), p. 157.

\bibitem{Littlewood} P. B. Littlewood and C. M. Varma, Phys. Rev.
B26 (1982) 4883. 
\end{thebibliography}
\end{document}